\documentstyle[12pt,procsla]{article}
  
  \catcode`\@=11
  \long\def\@makefntext#1{
  \protect\noindent \hbox to 3.2pt {\hskip-.9pt  
  $^{{\ninerm\@thefnmark}}$\hfil}#1\hfill}		
  
  \def\@makefnmark{\hbox to 0pt{$^{\@thefnmark}$\hss}}  
  	
  \def\ps@myheadings{\let\@mkboth\@gobbletwo
  \def\@oddhead{\hbox{}
  \rightmark\hfil\ninerm\thepage}   
  \def\@oddfoot{}\def\@evenhead{\ninerm\thepage\hfil
  \leftmark\hbox{}}\def\@evenfoot{}
  \def\sectionmark##1{}\def\subsectionmark##1{}}
  
\begin{document}
  
  \centerline{\normalsize\bf REACTOR NEUTRINO PHYSICS - AN UPDATE
\footnote
{Invited talk at the $8^{th}$ Int. Workshop on ''Neutrino Telescopes'',
Venice, February 1999.}}

  \baselineskip=16pt
  
  \centerline{\footnotesize FELIX BOEHM}
  \baselineskip=13pt
  \centerline{\footnotesize\it Physics Department, California Institute
  of Technology,
  }
  \baselineskip=12pt
  \centerline{\footnotesize\it Pasadena, CA 91125, USA}
  \centerline{\footnotesize E-mail: boehm@caltech.edu}
  \vspace*{0.3cm}
  
  \vspace*{0.9cm}
  
\abstracts{We review the status and the results of reactor neutrino
experiments.  Long baseline oscillation experiments at Palo Verde and
Chooz have provided limits for the oscillation parameters while the
recently proposed Kamland experiment at a baseline of more than 100km
is now in the planning stage.  We also describe the status of neutrino
magnetic moment experiments at reactors.}
   
  \normalsize\baselineskip=15pt
  \setcounter{footnote}{0}
  \renewcommand{\thefootnote}{\alph{footnote}}
  \section{The Reactor Neutrino Spectrum}

Nuclear reactors provide high flux low energy neutrinos and are thus
well suited for exploring $\bar{\nu}_e$ disappearance for small values
of the mass parameter $\Delta m^2$. As the neutrino energy $E_{\nu}$
enters as the ratio $L/E_{\nu}$, $L$ being the distance between
reactor and detector, a reactor experiment with $E_{\nu}$ around 5 MeV
at $L$ = 1 km has the same sensitivity to $\Delta m^2$ as an
accelerator experiment with $E_{\nu}$ = 5 GeV at 1,000 km.

The reactor neutrino flux and the neutrino-proton cross sections are
depicted together with their product in Fig. 1\cite{vogel}.

  \section{Oscillation Experiments}

The results from atmospheric neutrino experiments, such as those from
Kamiokande\cite{kamio} have triggered reactor neutrino studies in the
parameter region $\Delta m^2$ between $10^{-2}$ and $10^{-3}$ $eV^2$.
Two experiments, both at $L$ around 1 km, have been conducted
recently, one at the French reactor at Chooz\cite{chooz} and the other
at the Palo Verde\cite{pv} site in Arizona, USA.  Both experiments now
have results and we shall describe them below. Another experiment, the
Kamland\cite{kamland} experiment, at a much larger distance, is still
in the proposal stage.  To illustrate the effect from oscillations on
the positron spectrum, we show in Fig. 2 the expected spectrum for
Chooz or Palo Verde for the case of no oscillations as well as for the
set of oscillation parameters favored by the Kamiokande.  Clearly, the
effects on the spectrum are quite pronounced.

The Chooz and the Palo Verde experiments are based on the inverse
neutron decay reaction, $\bar{\nu_e} + p = e^{+} + n$, and rely on a
measurement of both, the $e^{+}$ and the $n$ as a correlated
signature. Both experiments make use of Gd loaded liquid scintillator.
Gd loading reduces the capture time owing to its large thermal neutron
capture cross section, and also gives rise to a high energy gamma
cascade of up to 8 MeV. Both features are valuable, the short capture
time helps reduce random coincidences and the large gamma ray energy
allows reduction of backgrounds as the energy threshold can be set
above that of radioactive decay products. In both experiments the
amount of Gd dissolved in the scintillator is about 0.1\% by weight.
At a distance of ca 1 km from the reactor the detector response is
about 5 events per day per ton of scintillator. The Chooz experiment
takes advantage of an existing deep tunnel reducing the cosmic ray
muon background substantially, while the Palo Verde experiment being
in a shallow underground laboratory has to cope with a larger muon
rate.  Because of this difference in shielding, the two detectors had
to be designed differently. While the Chooz detector consists of a
homogeneous central volume of Gd scintillator, the Palo Verde detector
is made from finely segmented detector cells.

\section{The Palo Verde Experiment}

The Palo Verde experiment, a collaboration\cite{pv} between Caltech,
Stanford University, University of Alabama, and Arizona State
University, is installed near the Palo Verde nuclear power plant (3
reactors, 11 GW thermal power) in the desert to the west of Phoenix,
Arizona.  The detector is installed in an underground cave with 32 mwe
overhead at a distance of $L$ = 890 m from reactors 1 and 3, and 750 m
from reactor 2.  Each reactor is shut down for refueling for a period
of ca 40 days every year which allows us to obtain background data.

The detector, shown schematically in Fig. 3, has a fiducial volume of
12 tons.  Its liquid scintillator, whose composition is 60\% mineral
oil, 36\% pseudocumene, 4\% alcohol, and 0.1\% Gd, was developed in
collaboration with Bicron\cite{piepke}. It has an effective light
attenuation length of 10m for 440nm light. There are 66 cells, each 9m
long, of which 7.4m are active and 0.8m on each end serve as an oil
buffer.  There is a 5 inch low radioactivity $Electron Tubes$
photomultiplier attached to each end, allowing both, the anode and the
last dinode to be read out. A blue LED installed at 0.9m from each PM,
as well as optical fibers, allow each individual cell to be monitored.
A passive water shield, 1m thick, surrounds the block of active cells
to help shield against radioactivity as well as muon induced neutrons.
An active veto counter consisting of 32 12m long MACRO cells is placed
on all 4 long sides while a removable end-veto counter protects the
ends of the cell matrix.

A diagram of the detector response showing the $\bar{\nu}_e$ reaction
and the gamma rays from Gd capture is given in Fig. 4.   

A neutrino signal consists of a fast (30ns) $e^+ \gamma \gamma$
trigger within a block of 3 x 5 cells, with the first hit having $E
\geq 500 keV$, and the second hit $E \geq 30 keV$. This second hit
includes the Compton response from the 511 keV annihilation gammas.
This fast triple coincidence is followed by a slow (200 $\mu s$)
signal due to the 8 MeV gamma cascade following neutron capture in Gd
within a 5 x 7 scintillator cell matrix.
  
  \subsection{Calibrations}

Energy calibrations were carried out with the help of small sources
that were introduced through a set of Teflon tubes installed alongside
a group of detector cells. The sources were $^{228}$Th (2.6 MeV
$\gamma$) depicted in Fig. 5, $^{65}$Zn, $^{137}$Cs, and $^{57}$Co
(0.12 MeV).  Response from these sources at various positions allowed
us to determine and monitor the attenuation length of the
scintillator. We found only a negligible decline over the period
measured. Fig. 6 shows the light yield along the scintillator cell.

The PMT linearity was obtained with the help of a fiber optics flasher.
Single photo-electron peaks were monitored with a blue LED.

To obtain the positron efficiency of the detector, we used the
positron emitter $^{22}$Na. A calibrated $^{68}$Ge
source\cite{piepke1} dissolved in a special cell will be implemented
later in the experiment.  The neutron efficiency was obtained with the
help of a calibrated AmBe source.  This source had a strength of 150
n/s and was used in a tagged mode, i.e.  in coincidence with the
4.4MeV gamma from $^{12}$C*.  The average efficiency (over the
detector) was also calculated with the help of a Monte Carlo
simulation. For our cuts, the resulting efficiency was found to be
0.159. Combined with a data acquisition efficiency of 0.82 and a
veto-live efficiency at our veto rate of 2 kHz, we determined the
total efficiency for observing a neutrino to be 0.082.

We calculated the number of neutrino interactions in the detector for
the case of no-oscillations from the neutrino flux from the 3
reactors, the cross section and the number of protons in the detector,
and found (for a given moment in the fuel cycle) an interaction rate
of 201 $\pm$ 2 per day.

\subsection{Results from the first 72 days of data}

We now present an early data set consisting of 39 days (November and
December 1998) of data with 3-reactors-on and 33 days (October 1998)
of data with 2-reactors-on (reactors at 890m and 750m) corresponding
to a $\nu$ flux of 100\% and 71\%, respectively.  The event rates were
found to be 39.1 $\pm$ 1.0 (st) per day and 32.6 $\pm$ 1.0 per day,
respectively, with a difference of 6.4 $\pm$ 1.4 (st) attributed to
the 890m reactor.  The signal-to-background ratio was found to be 1.2.
Corrected for efficiency, the number of neutrino interactions from the
890m reactor was found to be 77 $\pm$ 17 (st) $\pm$ 11 (sys) per day.
This is to be compared with the calculated number for the 890m reactor
of 59 $\pm$ 2 per day.  The correlated positron spectra (fourfold
coincidences) are shown in Fig. 7 and the time structure for these
events from the data and from Monte Carlo is given in Fig. 8 from
where it can be seen that the effective neutron capture time in our
detector is 35$\mu$s.

In Fig. 9 we show the exclusion plot obtained from our 72 day data
run, together with the parameter range allowed from the Kamiokande
atmospheric results.  Our results rule out
$\nu_{\mu}\leftrightarrow\nu_e$ oscillations at the 90\% confidence
level, in agreement with the results from the Chooz experiment to be
discussed below.

\subsection{The neutron angular distribution}

The segmentation of our detector allows us to study the
neutrino-neutron angular correlation and to use it for independent
background determinations.  From kinematics we find that the neutron
moves preferentially in the direction of the incoming neutrino, with
an angular distribution \begin{equation} \cos (\theta_{\nu,n})_{max} =
[ 2\Delta /E_{\nu} - (\Delta^2 - m^2 )/E_{\nu}^2 ]^{1/2}\,,
\end{equation} where $\Delta = M_n - M_p$.

From Monte Carlo simulation it is found that the neutron scattering
largely preserves the angular distribution, resulting in a shift of
the mean coordinate of the neutron capture center $\langle x \rangle$
= 1.7 cm\cite{vogelbeacom}. The angular spread after scattering is
very pronounced as can be seen in Fig. 10.  It should be noted that
this effect was first studied by Zacek\cite{zacek} in connection with
the segmented Goesgen detector where the forward/backward ratio was
found to be as large as a factor of 2.

Results from our first run give an asymmetry expressed as events in
the half plane away from the reactor (forward) minus events in the
half plane toward the reactor (backward) of 109 $\pm$ 44, in agreement
with a Monte Carlo simulation.

\subsection{Plans for the future of the Palo Verde experiment}

We plan to continue data runs through 1999, a period which includes
two refueling cycles, giving us a tenfold larger statistical accuracy
and correspondingly larger sensitivity to the mixing angle. We shall
continue to make use of the forward-backward asymmetry and we shall
also implement a $^{68}$Ge run to determine the detector efficiency
with higher accuracy.

\section{The Chooz Experiment}

An experiment with similar aim, however with a somewhat different
detector was carried out at Chooz by a French-Italian-Russian-US
collaboration\cite{chooz} This experiment and its results have been
published, and we will review it here with just a few highlights.

The Chooz detector is comprised of three regions, a central region
containing 5 tons of Gd loaded liquid scintillator and surrounded by
an acrylic vessel, a containment region with 17 tons of ordinary
liquid scintillator, and an outer veto region with 90 tons of
scintillator.  \newpage Fig. 11 shows schematically the arrangement of
the Chooz detector.

The inner two regions are viewed by a set of photomultipliers.  An
independent set of PM detects the light from the veto region.  While
the positron response is obtained from a signal in the inner region,
the neutron response comprises signals from the inner region as well
as from the containment region, resulting in a well contained and well
resolved Gd capture sum peak at 8 MeV.  As mentioned earlier, the
Chooz detector is installed in a tunnel, thus reducing the correlated
background to less than 10\% of the signal.

The published data was obtained at various power levels of one of the
two Chooz reactors.  A total of 1320 neutrino events was accumulated
in 2718 live hours. Normalized to the full power of the two reactors
(8.5 GW th) the event rate corresponds to 25.5 $\pm$ 1.0 per day where
the error includes contributions from the reaction cross section, the
reactor power, and the number of protons in the target.  The ratio of
measured-to-expected neutrino signal is 0.98 $\pm$ 0.04 (st) $\pm$
0.04 (sys).  The reactor-off rate was 1.2 $\pm$ 0.3 per day.  The
total efficiency of the detector was found to be 0.71 $\pm$ 0.02.

The positron energy spectrum for reactor-on and reactor-off is shown
in Fig. 12, together with a plot of the ratio of
measured-to-calculated spectrum.

Fig. 13 shows the Chooz exclusion plot. Clearly, the Kamiokande region
is excluded with high confidence level, implying the absence of
$\nu_{\mu} \leftrightarrow \nu_e$ oscillations.  The mixing angle
limit for large $\Delta m^2$ from this analysis is $sin^2 {2\theta} <
0.18$ at 90\% CL. While an analysis based on the widely accepted
method by Feldman and Cousins\cite{feldman} gives $sin^2{2\theta} <
0.22$ (90\%).  The 90\% limit for $\Delta m^2$ for maximum mixing from
this experiment is 0.9 x $10^{-3} eV^2$.

Recently, the Chooz collaboration has compared the spectrum from
reactor 2 which is at $L$ = 998m to that of reactor 1 at $L$ = 1115m.
The relative spectra from the two reactors at different distances
provide information on oscillations which is independent of the
absolute yields. Their analysis leads to an exclusion plot consistent
with, however less stringent than that of Fig. 13. The relative
positron yield is $Y$ (1115m) / $Y$ (998m) = 0.96 $\pm$ 0.06 (st)
$\pm$ 0.015 (sys).

The Chooz collaboration has completed data taking and is now
finalizing the data analysis. Their analysis will also include a
discussion of the neutron angular distribution as mentioned in the
section above on Palo Verde.

\section{The Kamland Experiment}

The Kamland\cite{kamland} experiment described by A. Suzuki in the
following lecture will be the ultimate long baseline reactor
experiment, destined to explore $\bar{\nu}_e$ disappearance at very
small $\Delta m^2$. It will be sensitive to exploring the large mixing
angle solar MSW solution. The experiment will also address the small
mixing angle solar MSW at low neutrino energy, as well as, by invoking
seasonal variations, the "just so" solution.

The neutrinos originate from 16 nuclear power reactors at distances
between 100 km and 300 km from the Kamland detector. The detector will
be a 1 kT liquid scintillator to be installed in the former Kamiokande
cavity. With an expected event rate of 1075/y the sensitivity to
$\Delta m^2$ will be 4 x $10^{-6} eV^2$.

\section{Neutrino Magnetic Moment}

If neutrinos have mass, they may have a magnetic moment.  An
experimental effort to look for a neutrino magnetic moment, therefore,
is of great interest. Some indications for a magnetic moment have come
from the apparent correlation of the signal in the $^{37}$Cl
experiment with solar activity, suggesting a value of $10^{-11} -
10^{-10} \mu_{Bohr}$.  In addition, considerations of a possible
resonant spin flavor precession (RSFP), and also of neutrino
interactions in supernovae have been mentioned.

The neutrino magnetic moment contributes to the $\nu_e e $
scattering\cite{vogelengel} as shown in Fig. 14.  This contribution is
most pronounced at low electron-recoil energy. At about 300 keV the
magnetic moment scattering is roughly equal to the weak scattering.

Previous results by Reines et al.\cite{reines} from scattering reactor
neutrinos on electrons in a 16 kg plastic scintillator have given
$\mu_{\nu} = 2 - 4 \times 10^{-10} \mu_B$. More recently, Gurevitch et
al.\cite{gurevitch}, using a 103kg $C_6 F_6$ target have found
$\mu_{\nu} < 2.4 \times 10^{-10} \mu_B$, and Derbin et
al.\cite{derbin} with a 75kg Si target have found $\mu_{\nu} < 1.8
\times 10^{-10} \mu_B$.

An effort to obtain a value or stringent limit of $\mu_{\nu}$ is now
underway by the MUNU experiment, a
Grenoble-Munster-Neuchatel-Padova-Zurich collaboration\cite{amsler}
which has built a 1000 liter $CF_4$ TPC at 5atm (18.5kg), surrounded
by an anti-Compton shield. This detector is now installed at the Bugey
reactor in France. The expected event rate in the interval of 0.5 to 1
MeV recoil energy is 5.1 per day, at an expected background of 4.5 per
day.  Implementing the angular correlation of the scattered electrons
with respect to the incoming reactor neutrinos is expected to enhance
the signal-to-noise significantly.  A schematic view of the MUNU TPC
is shown in Fig. 15.

\section{Conclusion}

Reactor neutrinos with their low energies are well suited to explore
small $\Delta m^2$ for the $\bar{\nu_e}$ disappearance channel.  From
the results of the Chooz and Palo Verde experiments it can be
concluded that the atmospheric $\nu_{\mu}$ deficiency cannot be
attributed to $\bar{\nu}_{\mu} \leftrightarrow \bar{\nu}_e$
oscillations. The Chooz experiment has ruled out this channel with
large confidence level, and the first data set from Palo Verde
excludes it at 90\% CL. To improve on the mixing angle sensitivity in
these experiments so as to shed light on a possible 3-flavor solution
will be a challenging task.  The Kamland experiment at very large
baseline is now on the drawing board.  Searches for a neutrino
magnetic moment from the MUNU experiment are in progress.

\newpage

\section{Figure Captions}

\vspace*{0.5cm}
\fcaption
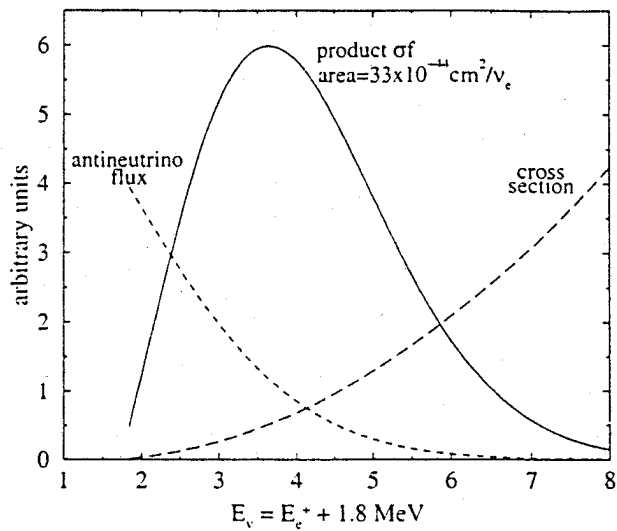
{Energy spectrum, cross section and yield of neutrinos from a reactor.}

\vspace*{0.5cm}
\fcaption
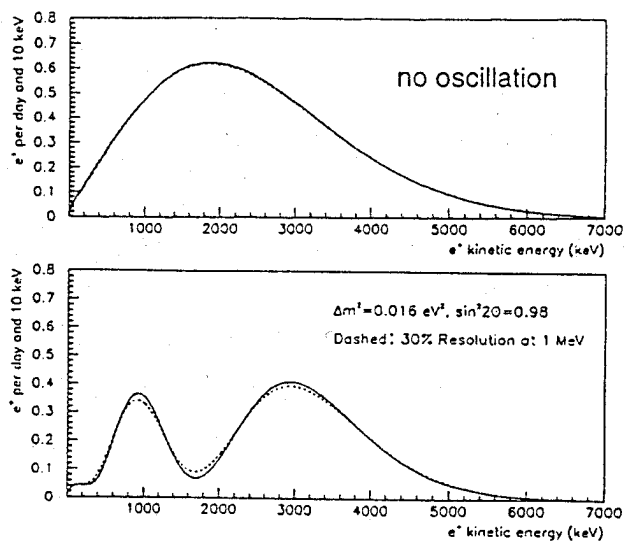
{Expected positron spectra for the Chooz or Palo Verde
experiment for "no oscillations" and for oscillations given by the
Kamiokande parameters.} 

\vspace*{0.5cm}
\fcaption
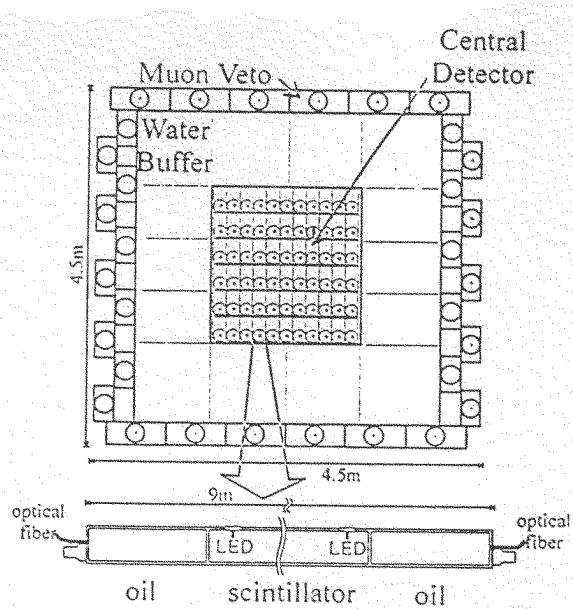
{Schematic view of the Palo Verde detector.}

\vspace*{0.5cm}
\fcaption
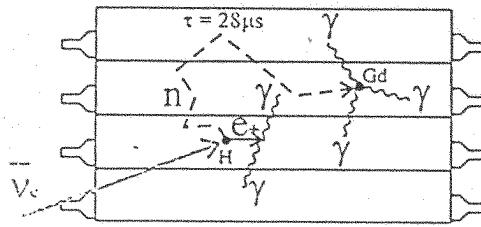
{Illustration of the neutrino reaction in the 
matrix of Gd loaded scintillator.}

\vspace*{0.5cm}
\fcaption
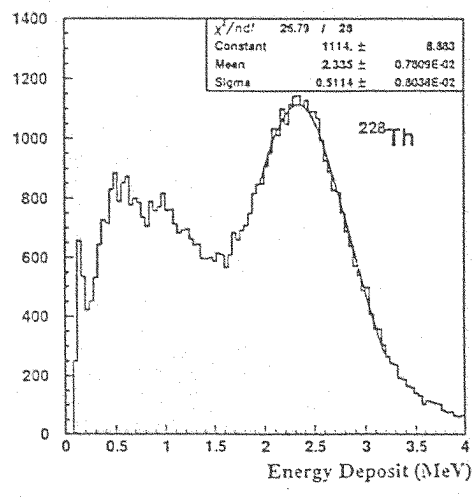
{Energy spectrum from a $^{228}$Th source ($E_\gamma$ = 2.6 MeV).}

\vspace*{0.5cm}
\fcaption
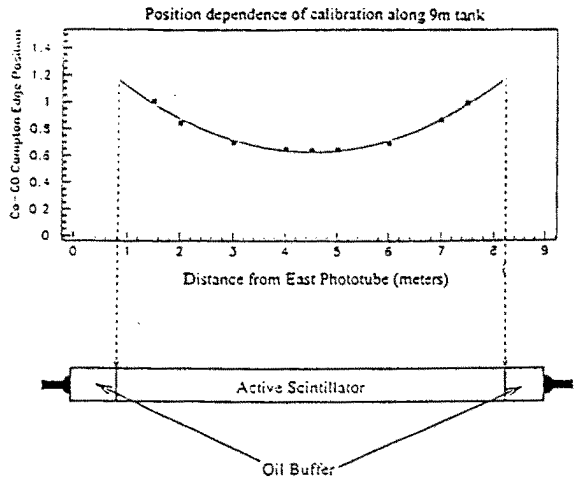
{Light yield along scintillator cell. The attenuation length of
the Gd scintillator is 10m.}

\vspace*{0.5cm}
\fcaption
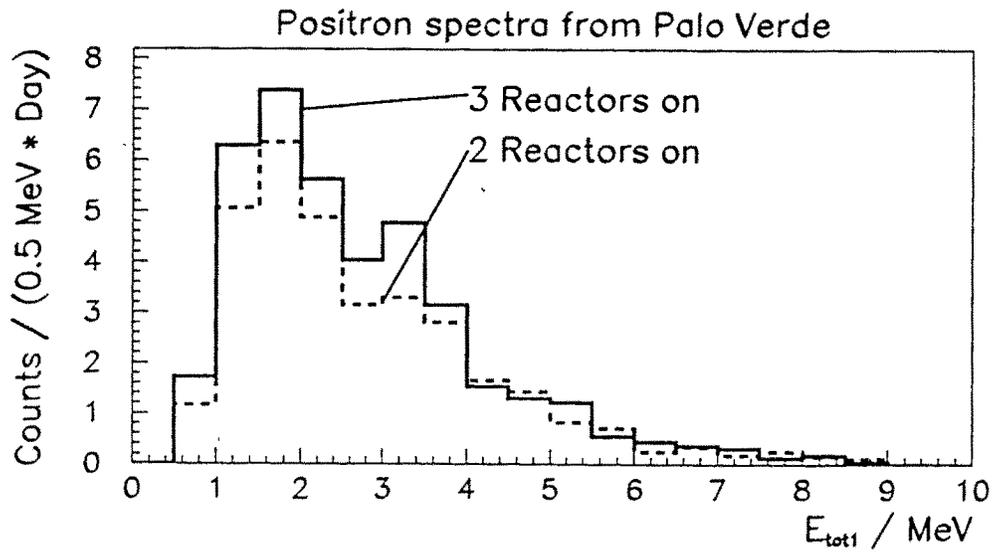
{Correlated positron spectrum from a 3-reactor run (100\% $\nu$ flux) 
and a 2-reactor run (71\% flux).}

\vspace*{0.5cm}
\fcaption
{Decay time of the fourfold coincidence gives the neutron capture
time in our Gd-scintillator.}

\vspace{0.5cm}
\fcaption
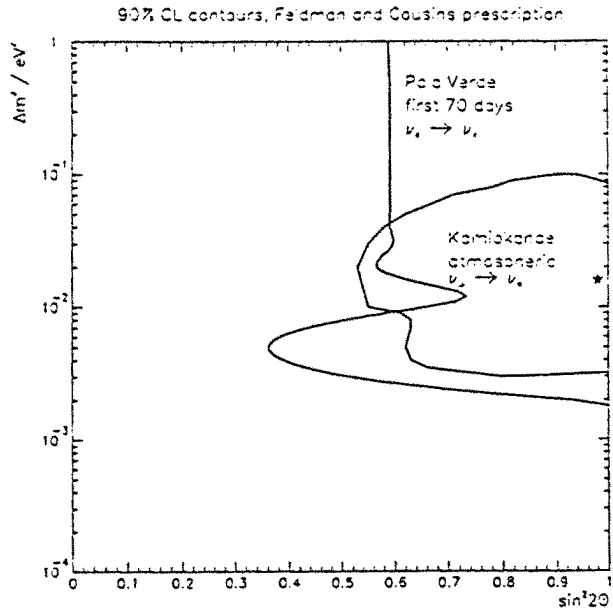
{Region in the parameter space excluded at 90\% CL for our 72 day run.
Most of the Kamiokande allowed region can be excluded from this data.}

\vspace*{0.5cm}
\fcaption
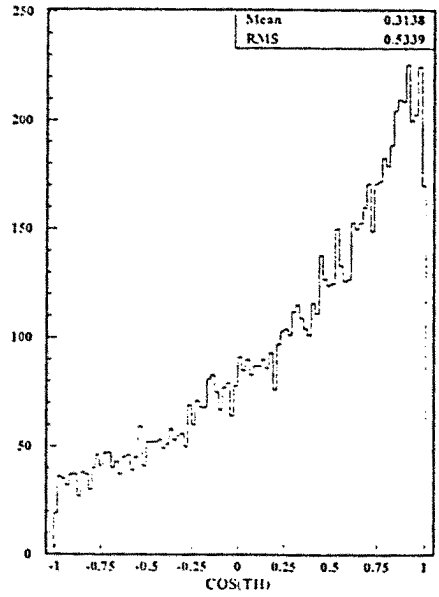
{Angular distribution of scattered (moderated) neutrons with 
regards to the neutrino direction.}

\vspace{0.5cm}
\fcaption
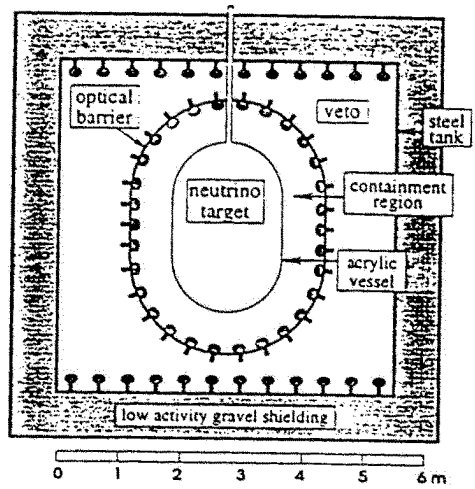
{Schematic arrangement of the Chooz detector.}

\vspace{0.5cm}
\fcaption
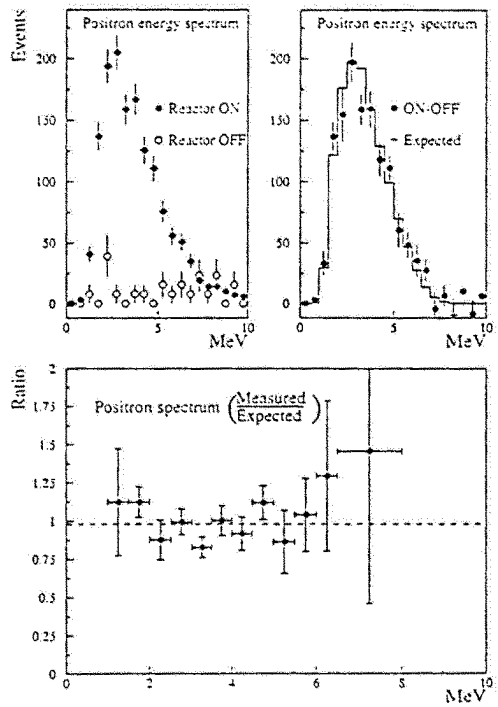
{Positron energy spectra from the Chooz experiment.}

\vspace*{0.5cm}
\fcaption
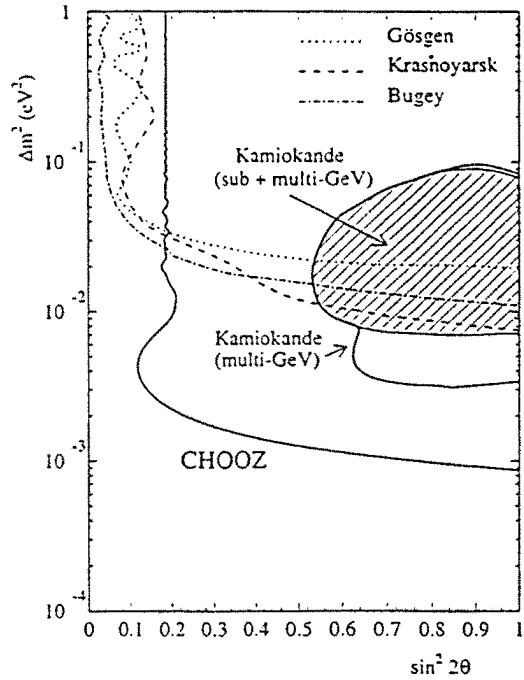
{Exclusion plot from the Chooz experiment.}

\vspace*{0.5cm}
\fcaption

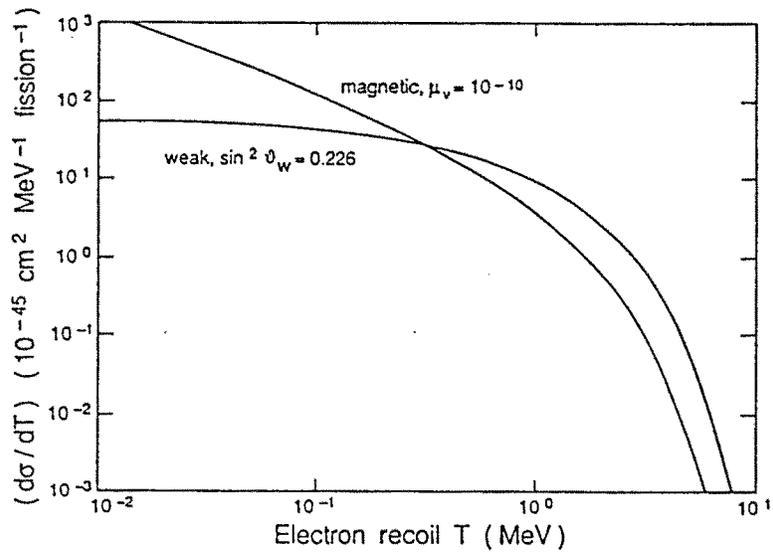
{Contribution of the neutrino magnetic moment to the $\bar{\nu}_e e
\rightarrow \bar{\nu}_e e $ scattering, averaged over the reactor
$\bar{\nu}_e$ spectrum. The purely weak cross section is also shown.
(From Vogel and Engel\cite{vogelengel})}

\vspace*{0.5cm}
\fcaption
{Layout of the MUNU detector for the measurement of the neutrino
magnetic moment.}
  
  
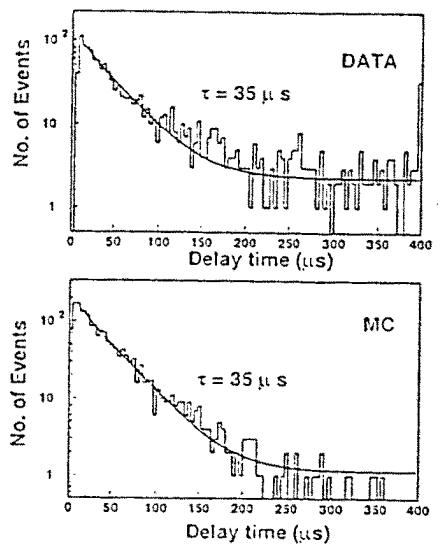
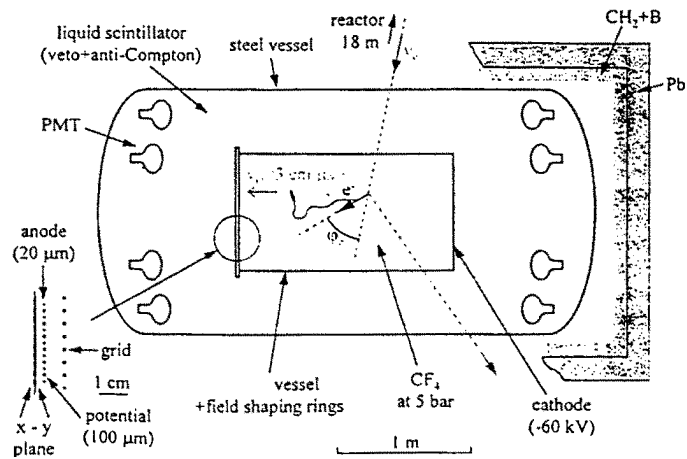
\end{document}